\documentclass[11pt]{article}
\usepackage[margin=1in]{geometry}
\usepackage{graphicx, wrapfig} 
\usepackage[inkscapelatex=false]{svg}
\usepackage[export]{adjustbox} 

\usepackage{pifont, csquotes}
\usepackage{hyperref, lineno} 

\usepackage{chemformula, braket,siunitx}
\usepackage[version=4]{mhchem}
\usepackage{soul}  

\usepackage[style=numeric, sorting=none]{biblatex}
\usepackage{mathtools}
\usepackage{caption, subcaption, authblk}
\usepackage{booktabs}

\usepackage{accents}
\usepackage{float}
\usepackage[ruled,vlined]{algorithm2e}

\title{High nucleotide skew palindromic DNA sequences function as replication origins due to their unzipping propensity}
\author{Parthasarathi Sahu}
\author{Sashikanta Barik}
\author{Koushik Ghosh}
\author{Hemachander Subramanian\thanks{Corresponding author: hsubramanian.phy@nitdgp.ac.in}}

\affil{Department of Physics, National Institute of Technology Durgapur, India}

\begin{document}

\maketitle

\begin{abstract}
 \textit{Locations of DNA replication initiation in prokaryotes, called \enquote{origins of replication}, are well-characterized. However, a mechanistic understanding of the sequence-dependence of the local unzipping of double-stranded DNA, the first step towards replication initiation, is lacking. Here, utilizing a Markov chain model that was created to address the directional nature of DNA unzipping and replication, we model the sequence dependence of local melting of double-stranded linear DNA segments. We show that generalized palindromic sequences with high nucleotide skews have a low kinetic barrier for local melting near melting temperatures. This allows for such sequences to function as replication origins. We support our claim with evidence for high-skew palindromic sequences within the replication origins of mitochondrial DNA, bacteria, archaea and plasmids.}     
\end{abstract}

\subsubsection*{Keywords}
Asymmetric cooperativity, kinetic asymmetry, Origin of Replication, sequence-dependent kinetics, GC-skew, RY-skew, bubble formation, DNA unzipping

\section*{Statements and Declarations}

\subsection*{Competing interests}
The authors declare no competing interests.

\subsection*{Acknowledgments}
Support for this work was provided by the Science \& Engineering Research Board (SERB), Department of Science and Technology (DST), India, through a Core Research Grant with file no. CRG/2020/003555 and a MATRICS grant with file no. MTR/2022/000086. 

\subsection*{Data Availability}
The sequences and algorithms used in the analysis are available at: 

\noindent
https://github.com/ParthaTbio/UnzippingDNA

\newpage
\section*{Introduction}

DNA stores information in its sequence. Apart from encoding information about amino acid sequences that get translated into proteins, DNA also encodes information about its own replication, for example, through sequences that are recognized by DnaA binding protein in prokaryotes \cite{pabo1984protein, fuller1983purified, leonard2013dna} and by origin recognition complexes (ORC) in eukaryotes \cite{bell1992atp, lee1997architecture, parker2017mechanisms}. What role does DNA play in such recognition processes? Does it actively participate in directing ORC towards specific locations on the genome or is it simply a passive template waiting to be traversed and recognized by ORC? 
\par
 Here we theoretically show that thermally-activated \cite{chen1992energy}, \textit{rapid, cooperative unzipping} of specific sequences at replication origins, mediated through favorable sequence-dependent \textit{kinetic} interactions between neighboring base pairs, results in origin activation. The proposed involvement of sequence-dependent unzipping \textit{kinetics} in origin activation is in contrast to suggestions of \textit{thermodynamics}-dictated origin functionality proposed earlier \cite{lee2023and, prioleau2016dna, chua2012mechanics, mechali2001dna}. Utilizing a \textit{kinetics-based} model proposed earlier to understand the evolutionary origins of strand directional asymmetry in DNA, we demonstrate that, near the double-stranded DNA (dsDNA) melting temperature, RY-palindromic (or MK-palindromic) sequences with high nucleotide skews have low \textit{kinetic} barriers for local, cooperative unzipping, thereby instantiating replication origins. By RY-palindrome, we mean a DNA sequence expressed in terms of purines ($R=\{G,A\}$) and pyrimidines ($Y=\{C,T\}$), instead of the usual four nucleotides, being palindromic; i.e, a sequence written in RY alphabet being the same as its reverse complement. For example, the sequence $5'\text{-YYRYYYRRRYRR-}3'$ is an RY-palindrome, since its reverse complement is the same sequence. MK-palindromic sequences can be similarly defined, where amino-keto grouping ($K=\{G,T\}$ and $M=\{C,A\}$) of the nucleotides is used, instead of purine-pyrimidine grouping. Clearly, conventionally defined palindromic sequences are a subset of RY-palindromic and MK-palindromic sequences. 
 
 \par
 By the term \enquote{nucleotide skew}, we imply an excess of one type of nucleotide (say, purine) over its base-pairing complement (pyrimidine) on a specific single strand. We quantify it by calculating the difference between the number of purines and the number of pyrimidines cumulatively, beginning at the $5'$-end and summing over increasing lengths from zero to the entire length of the sequence, and plotting the skew as a function of the summed length on the genome \cite{grigoriev1998analyzing, tillier2000contributions, mclean1998base}. Mathematically, the evaluation of the RY cumulative skew, $W_{RY}$, can be expressed as
 \begin{equation}\label{cumsum}
    W_{RY}(t) = \sum_{i=1}^{t} (\delta_{S(i),R} - \delta_{S(i),Y}), 
\end{equation}
where, S is the genomic sequence of length $N$ with each element being any one of the four nucleotides, $R=\{G,A\}$, $Y=\{C,T\}$, and $t=1\ldots N$. Amino-Keto or MK cumulative skew $W_{MK}$ can be similarly defined, with $K=\{G,T\}$ and $M=\{C,A\}$. A high nucleotide skew RY-palindrome implies sequences such as $5'\text{-YYYYYYRRRRRR-}3'$, where the six nucleotides at $5'$-end are highly skewed towards pyrimidines and the nucleotides at $3'$-end are entirely purines. This skew analysis is commonly employed to find origins of replication in both prokaryotes and eukaryotes, over a length scale of megabases \cite{tillier2000contributions, frank2000oriloc, rocha2004replication, niu2003strand, dai2005dna, marsolier2012asymmetry, bartholdy2015allele}. At the outset, we would like to clarify that, by our usage of sequences such as $5'\text{-YYYRRR-}3'$, we imply either $5'\text{-TTTAAA-}3'$ or $5'\text{-CCCGGG-}3'$, and never a mixture of both, such as $5'\text{-CTCGAG-}3'$. Such a segregation of pure GC and pure AT sequences is done in order to isolate the effect of kinetics on unzipping of the sequences, without the meddling influence of thermodynamics. A more nuanced reason for this choice is provided at the end of the Results section. 

\par
We have qualitatively argued in our earlier paper \cite{sequence_dependent} that the \textit{kinetics} of unzipping of DNA, which determines the rate of self-replication at temperatures below the melting point of the inter-strand hydrogen bonds, is determined by the sequence of the DNA, due to the presence of a property we call \enquote{\textit{asymmetric cooperativity}}. In this article, we \textit{quantitatively} demonstrate the effect of the DNA sequence on the \textit{kinetics} of unzipping of DNA double strands, and show that near melting temperature of DNA, RY- or MK-palindromic, high-nucleotide-skew sequences unzip faster than other sequences, and hence, are found near replication origins. Since unzipping is the rate-limiting step during DNA replication below the melting temperatures of the double-strand, unzipping rates dictate replication rates, and hence the sequences' evolutionary superiority. Below, we first introduce our model assumptions, elaborate on its mathematical implementation, and use the model to calculate and compare the unzipping times of various types of sequences.

\section*{Asymmetric Cooperativity Model}

In our earlier papers \cite{sequence_dependent, symmetry_breaking}, we have proposed the existence of \enquote{asymmetric cooperativity}, \textit{a kinetic property}, in DNA and its evolutionary progenitors, in order to rationalize the existence of certain evolutionarily counter-intuitive properties of DNA, such as its unidirectional strand construction and anti-parallel strand orientation. Both these properties together lead to the complicated lagging strand replication mechanism of DNA, involving piecemeal lagging strand construction and their eventual ligation. This complex self-replication process could have been avoided by utilizing a parallel-stranded heteropolymer, capable of bidirectional replication along a single strand template, for information storage. We have shown \cite{sequence_dependent, symmetry_breaking} that the reason for evolutionary selection of unidirectional, anti-parallel heteropolymer, such as DNA, as opposed to a bidirectional, parallel heteropolymer, is the evolutionarily advantageous \textit{sequence-dependent kinetics}: Due to the presence of asymmetric cooperativity, sequences dictate their own unzipping rates, and hence replication rates, thereby kickstarting evolutionary competition among themselves for resources such as monomers and energy supply. In this article, we demonstrate such evolutionary competition among different sequences by calculating the unzipping rates of these sequences and show that RY-palindromic, high-skew sequences emerge as winners of the competition, within a broad region of phase space defined by temperature and sequence length. Below, we introduce the property of asymmetric cooperativity and explain its influence on the unzipping kinetics of DNA double strands. It has to be strongly emphasized that the \textit{central premise} of this article is the presence of asymmetric cooperativity in DNA, from which the results will be shown to follow.

\begin{figure}[H]
     \centering
         \centering
         \includegraphics[width = \textwidth]{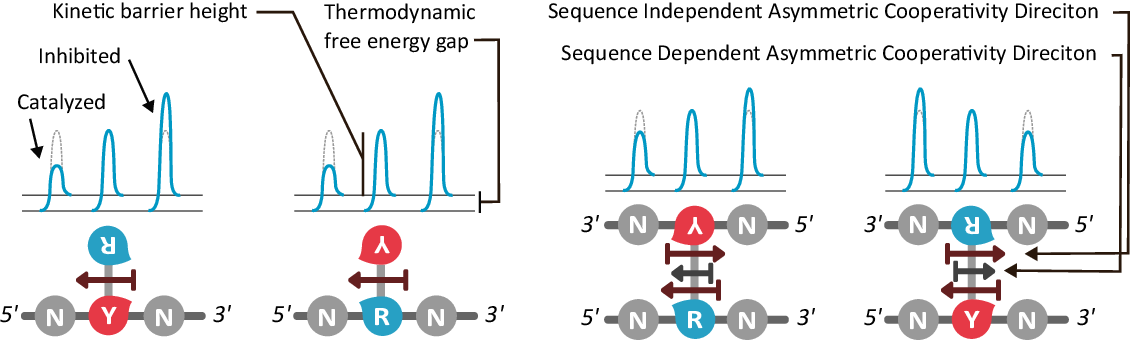}
         
        \caption{\textit{Illustration of the effect of one hydrogen bond on the kinetic barriers for hydrogen bond formation or dissociation of its left and right neighbor according to our model. (a) A hydrogen bond between a single nucleotide and a template strand reduces the kinetic barrier for new hydrogen bond formation towards the $5'$-end of the template, as represented by a brown arrowhead, and increases the kinetic barrier for new bond formation towards the $3'$-end, as represented by the bar-head. A barrier diagram is included above the diagram for illustration of barrier height changes due to asymmetric cooperativity. This asymmetric cooperative effect leads to unidirectional replication of the template strand. We term this sequence-independent asymmetric cooperativity. (b) When the hydrogen bond is between nucleotides that constitute fully formed double strands of DNA, the sequence-independent asymmetric cooperativities of the two strands cancel, due to their anti-parallel directions, and a remnant asymmetric cooperativity, weaker than the one in (a), remains to influence the kinetic barriers of neighboring hydrogen bonds to the left and right. This is represented as the smaller black arrow between the two larger brown arrows in (b) above. The direction of kinetic influence is dictated by the sequence, where $5'\text{-Y-}3'/3'\text{-R-}5'$ reduces the kinetic barrier of the right neighbor and increases the barrier of the left neighbor. The $180^0$-rotated $5'\text{-R-}3'/3'\text{-Y-}5'$ reduces the barrier to its left and increases it to the right. We term this sequence-dependent asymmetric cooperativity. }}
        \label{HemaFig1}
\end{figure}

\par
Within the asymmetric cooperativity model, a hydrogen-bonded base-pair on DNA \textit{kinetically} influences its left and right neighboring bonds asymmetrically and non-reciprocally. This implies that a hydrogen-bonded base-pair on a single-stranded DNA reduces the kinetic barrier for its left (or right) neighboring hydrogen bond formation or dissociation, and increases the kinetic barrier for its right (or left) bond formation or dissociation. This asymmetric influence of a hydrogen-bonded base pair on its immediate neighbors leads to unidirectional daughter strand construction on a single-stranded template, by reducing the barrier for new hydrogen bond formation at the strand-growth front, and increasing the barrier for already-formed base pairs behind the growth front, which improves intra-strand covalent bond formation probability. Therefore, the presence of asymmetric cooperativity improves the speed of daughter strand construction and could have led to DNA adopting such asymmetry, implementing it structurally as $3'\rightarrow 5'$ strand directionality \cite{symmetry_breaking}, leading to unidirectional strand construction in DNA. We call the asymmetric cooperativity on DNA \textit{single} strands as \enquote{sequence-independent} asymmetric cooperativity. 

In DNA \textit{double} strands, the directionalities of the two strands are opposite, since the strands are anti-parallel, and hence, sequence-independent asymmetric cooperativity of the two strands stands canceled, due to the opposing kinetic influence from both strands. If the base-pairs in DNA are formed between the same type of nucleotides, say, between two A's, such cancellation of asymmetric cooperativity from both strands would be complete, and there will be no asymmetric influence of neighboring hydrogen bonds, simply due to the left-right symmetry of the system ($5'\text{-A-}3'/3'\text{-A-}5'$ is left-right symmetric, which can be verified by rotating the base-pair). However, base-pairs form between distinct nucleotides, i.e., between A and T, and G and C, which are obviously left-right asymmetric ($5'\text{-T-}3'/3'\text{-A-}5'$ is left-right asymmetric, since rotating this base-pair leads to a different configuration, $5'\text{-A-}3'/3'\text{-T-}5'$), and can instantiate asymmetric cooperativity. Therefore, the cancellation of asymmetric cooperativity between the two anti-parallel strands is incomplete, due to heteromolecular base-pairing, leaving a remnant of asymmetric cooperativity, which is dependent on the \textit{orientation} of the base-pair, which distinguishes, say, $5'\text{-T-}3'/3'\text{-A-}5'$ from $5'\text{-A-}3'/3'\text{-T-}5'$. We call this base-pair orientation-dependent asymmetric cooperativity as \enquote{sequence-dependent} asymmetric cooperativity and posit that it is weaker than sequence-independent asymmetric cooperativity, in order to align the predictions of our model with fundamental observations regarding DNA replication, such as the lagging strand replication mechanism \cite {sequence_dependent}. Here, we assume that the sequence-dependent asymmetric cooperativity parameters to be the same for both $5'\text{-G-}3'/3'\text{-C-}5'$ and $5'\text{-A-}3'/3'\text{-T-}5'$ base pairs, thereby grouping C and T (pyrimidines), and G and A (purines), together. However, the direction of sequence-dependent cooperativity of AT base pair can vary across organisms even within a single biological domain \cite{grigoriev1998analyzing, perna1995patterns, charneski2011atypical}, with the direction of sequence-dependent asymmetric cooperativity of $5'\text{-A-}3'/3'\text{-T-}5'$ being opposite to that of $5'\text{-G-}3'/3'\text{-C-}5'$ in some organisms, leading to grouping of C and A (amino), and G and T (keto). The foregoing has been elaborated in more detail in our earlier papers \cite{symmetry_breaking, sequence_dependent}, and the reader is requested to refer to them for a thorough exposition of the asymmetric cooperativity model's explanations and experimental support.

\section*{Methods}
The goal of this paper is to quantitatively identify the fastest unzipping \textit{linear} sequence(s) in a given environment, among all possible sequences of the same length, and gain an understanding of its unzipping dynamics. We utilize the Continuous Time Markov Chain methodology to evaluate the unzipping times of various sequences \cite{gillespie1991markov, van1992stochastic}. We identify all possible configurations traversed by the fully-zipped double-stranded DNA towards its complete unzipping, and sample and store them in a vector $S$, termed state space or configurational space. These states correspond to all possible combinations of bonded/unbonded hydrogen bonds between the two strands of DNA. For example, the state $00001$ corresponds to all the hydrogen bonds between the two five-nucleotide-long strands broken, except the rightmost one. The state space size, therefore, will be $2^n$, where $n$ is the sequence length. We assume that transitions between two states can happen only if the two states differ in the status of a single hydrogen bond, thereby allowing only a single inter-strand hydrogen bond to form or break during a single transition. 

\begin{figure}[H]
    \centering
         \includegraphics[width = 1\textwidth]{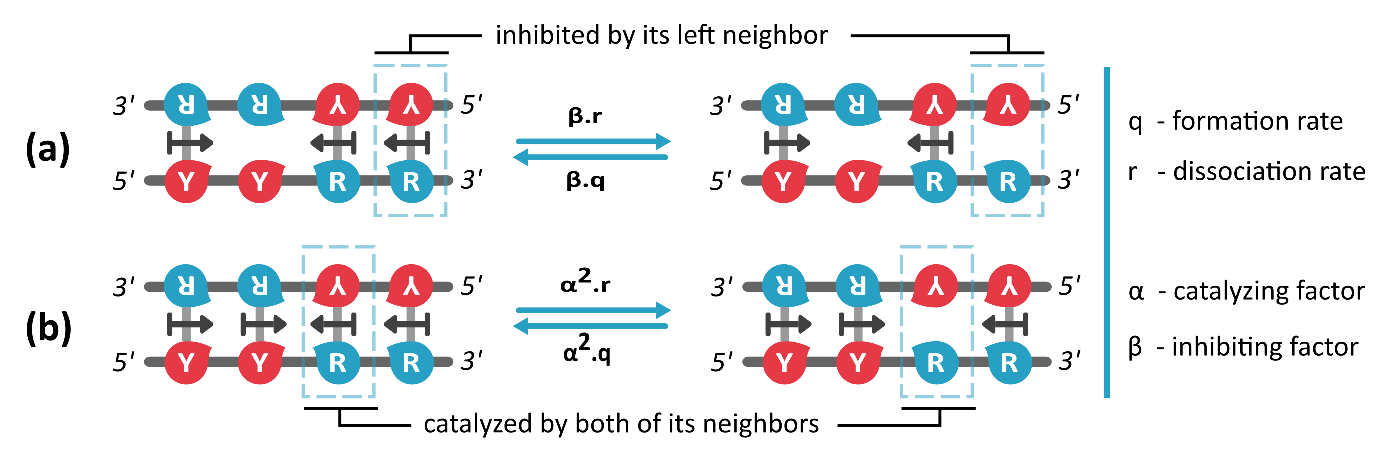}
    \caption{\textit{Examples of calculation of elements of rate matrix of the continuous-time Markov chain, constructed to evaluate DNA unzipping time. (a) The calculation of the rate matrix element for the transition between the states $1011$ and $1010$ is shown. The forward transition from $1011$ to $1010$ involves the dissociation of the rightmost H-bond. Since the bond's kinetic barrier is raised by the presence of a left-neighboring H-Bond in the $5'\text{-R-}3'/3'\text{-Y-}5'$ orientation, which reduces the barrier to its right, the rate of H-bond dissociation, $r$, is modified by an inhibiting factor $\beta$, resulting in an overall rate of $r\beta$. Similarly, the reverse transition from the state $1010$ to $1011$, denoting the formation of the rightmost H-bond, has a rate of $q\beta$, where $q$ is the formation rate of a single H-bond without any neighbors. (b) Rate matrix evaluation between the states $1111$ and $1101$. The dissociation rate of the third H-bond from the left, without any neighborhood influence, is $r$, which is modified due to the presence of both left and right neighbors. The orientation of the third H-bond's left and right neighbors is such that the kinetic barrier for the H-bond's dissociation is doubly reduced, thereby resulting in a dissociation rate of $r\alpha^2$, where $\alpha$ quantifies the catalyzing effect of each of the neighbors. The reverse transition, from $1101$ and $1111$, is then $q\alpha^2$. It has to be noted that $\beta$ and $\alpha$ alter both the formation and dissociation rates $q$ and $r$ equally, and hence are purely kinetic factors that leave the thermodynamics unaffected.}}
    \label{fig: rates}
\end{figure}

The transition rates $K_{ij}$ between any two states $i$ and $j$ in the state space $S$ are calculated as follows. The rate of formation and dissociation of an inter-strand H-bond, in the absence of any neighboring H-bonds, are denoted as $q$ and $r$ respectively. In the presence of a single neighboring H-bond (left or right), we take the cooperative effect into account by modulating $q$ and $r$ by a sequence-dependent asymmetric cooperativity factor of either $\alpha$ ($>1$, catalytic) or $\beta$ ($<1$, inhibitory), depending on the orientation of the neighboring base-pair. The dependence of the barrier height for the formation/dissociation of an H-bond on the orientation of its neighboring base-pair to its left or right is illustrated in Fig. \ref{fig: rates}. When both the left and right neighboring H-bonds are present, the bonding/unbonding rates are doubly modulated with appropriate parameters, depending on the orientation of both base pairs. Since both the strands of DNA are intact during unzipping, sequence-independent asymmetric cooperativity stands canceled due to the anti-parallel directions of both the strands and hence is not included in the model. The diagonal entries of the transition matrix $K$ are set to $K_{ii} = -\sum\limits_{\substack{i, i \neq j}} K_{ij}$ \cite{van1992stochastic}. The Mean First Passage Time is defined as the average time required for the Markov chain to reach the target state, which in our case is the fully unzipped DNA state, starting from the initial state, which is the fully zipped double-stranded state. This is calculated by inverting a modified rate matrix $K'$, obtained by eliminating the target state row(s) and column(s), as shown in the following equation \cite{inversemethod,master_equation,inversemethod1}:
\begin{equation}\label{direct inverse method}
    \textbf{$K^\prime \cdot{T} = -1$}
\end{equation}
The elements of the residence time matrix $T$, $T_{ij}$, provide the amount of time spent in state $j$, when the chain starts at state $i$, during its sojourn towards the target state. 

\par
 
\subsection*{Spectral decomposition} 
 
\par
As an alternative to equation (\ref{direct inverse method}), the hitting time can be calculated using eq (\ref{spectral terms}), which involves an eigen-decomposition of transition rate matrix as 
$K = \sum_l \lambda _l \ket{\phi}_l \bra{\psi}_l$, where, $\ket{\phi}_l$ and $\ket{\psi}_l$ are the left and right eigenvectors of $K$ corresponding to each eigenvalue $\lambda _l$ ($l = 0,1,2,... L-1$) \cite{buchete2008coarse, kells2020correlation}. The normalized left eigenvector $(\ket{\psi} _0)$ corresponding to zero eigenvalue $(\lambda_0)$ gives the steady state probability distribution $P_s$ of the Markov chain. Hitting time to transit from $i^{th}$ to $j^{th}$ state is given as
\begin{equation}\label{spectral terms}
    t_{ij} = \frac{1}{(P_s)_j}\sum_{l>0} \frac{1}{|\lambda_l|}\psi^l_j(\phi^l_j - \phi^l_i )
\end{equation}
Equation (\ref{spectral terms}) provides an alternate approach to study the unzipping process by decomposing the hitting time into $L$ different modes, where the minimum non-zero $|\lambda|$ value corresponds to the slowest mode of transition \cite{kells2020correlation}. We studied this mode to analyze the slowest transitions in the unzipping process.

When the rate matrix is not symmetric (i.e., Markov chain process is not at equilibrium), the eigenvectors lack orthogonality, hence $K$ may be symmetrized prior to the use of eq (\ref{spectral terms}) \cite{buchete2008coarse}. The symmetrized rate matrix corresponding to $K$ is given as $K_{sym} = P_s^{1/2} K P_s^{-1/2}$. New eigenvectors are calculated as; $\phi_{new} = P_s^{-1/2} \psi _{sym}$ and $\psi_{new} = P_s^{1/2} \psi _{sym}$ where $\psi_{sym}$ is the right eigenvector of $K_{sym}$. The new eigenvalues $\phi_{new}$ and $\psi_{new}$ can be plugged into eq (\ref{spectral terms}) in place of $\phi$ and $\psi$ for the decomposition of hitting time.

\subsection*{Temporal evolution of state probabilities}
We calculate the temporal evolution of probabilities of the Markov chain being in various states using the master equation $dP/dt = PK$ \cite{buchete2008coarse}. Here $P(t)$ is a vector whose elements are probabilities of occupation of all the states at time $t$. The formal solution to the above equation is
\begin{equation}\label{evolution of probability}
    P(t) = P(0)e^{Kt} 
\end{equation}
In our case, the Markov chain is assumed to begin from the fully zipped dsDNA state, whose probability at time $t=0$ is, therefore, $1$. The system eventually attains a steady state (Fig. \ref{fig: time evolution of probability}) when $P_se^{Kt} = P_s$, where $P_s$ is the steady-state probability distribution over all possible states \cite{kells2020correlation}.

\subsection*{Sequence Data Analysis} 
To validate our hypothesis that initiation of unzipping at replication origin requires asymmetric nucleotide distribution, we analyzed OriC sequences from prokaryotic domains / cellular components (mitochondria, bacteria, archaea and plasmids) to find potential skew in the nucleotide distribution within these sequences. Such distributions are typically visualized by GC-, RY- or MK-cumulative skew plots, where an upward trend in the curve indicates an abundance of G/R/K and a downward trend indicates C/Y/M abundance in the sequence \cite{grigoriev1998analyzing, dong2023doric}. Averaging the cumulative skew plots of a large number of sequences can help us observe the collective enrichment of one type of nucleotide group over another at specific locations in the genomes of a given biological domain / cellular component. 

In averaging the cumulative plots over all the organisms within a given domain / cellular component, we encounter two issues: I) The lengths of origin sequences are variable within a domain, and across domains, and hence need to be standardized. II) Within a domain, the identification of replication origin of a given organism relies on either RY or MK-cumulative skew \cite{dong2023doric, sernova2008identification}. With no information provided in the database regarding which of the two skews are used to identify origins, we need the algorithm itself to decide on the appropriate skew option. 

Solution for problem I: The length of mitochondrial origins is of the order of tens of base pairs, whereas, that of bacteria can range from hundreds to thousands of base pairs. Our analysis must be applicable for both these length scales. Moreover, for longer-length sequences, we must look for sequence signatures at the largest possible length scale, ignoring finer scale details that may obscure the big picture. We choose wavelet transform as our tool, since it is a good fit for our requirements of scale-independent analysis. Wavelet transforms allow us to compute large-scale information in a sequence, stored in \enquote{approximate coefficients}, without losing finer-scale information, which is stored in \enquote{detailed coefficients}. We standardized all skew plots, using wavelet transform, to a uniform length of 16 (8 for mitochondria) while preserving the global nature of the skew plot. In this transformation, the sequences are first trimmed on both ends to a length of nearest $2^n, n\in \{5,6,7...\}$, then a multi-level wavelet transform of the cumulative skew sequence brings the length down to 16 (or 8). Subsequently, these transformed skew plots are averaged over all organisms within each domain / cellular component to obtain four comprehensive cumulative skew plots representing each of the four domains / cellular components. 

\begin{figure}[H]
\centering
\includegraphics[width=1\textwidth]{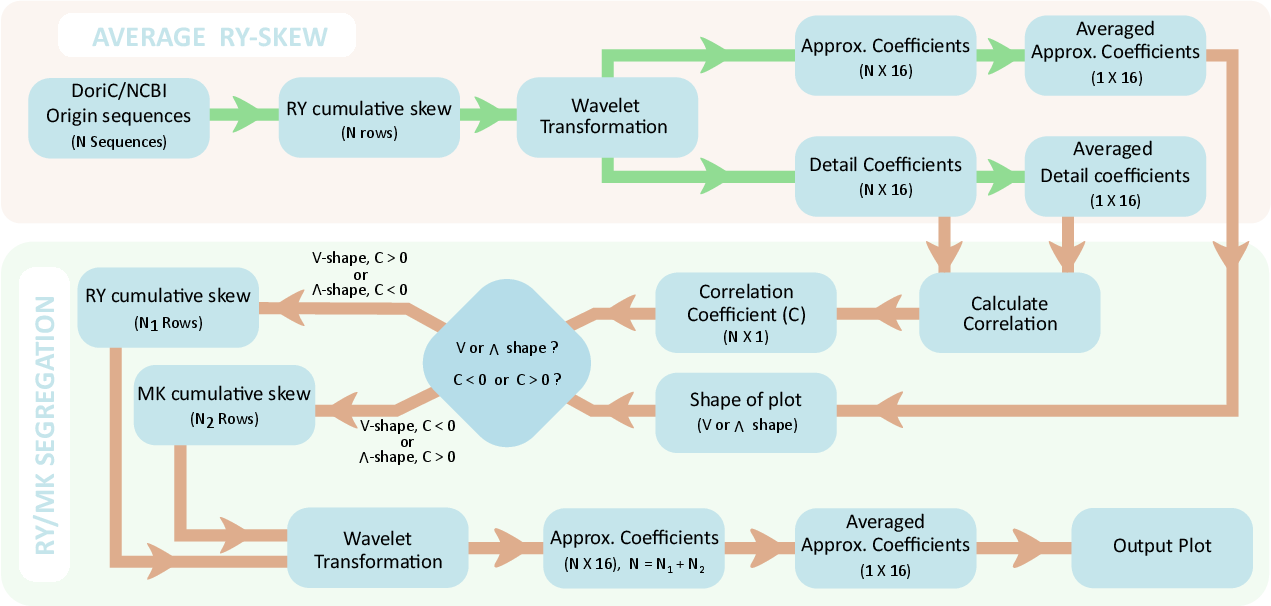}

    \caption{\textit{Algorithm for length standardization of sequences and RY/MK skew discrimination. This algorithm is applied for each of the four domains / cellular components. In the block titled `Average RY-skew', the cumulative skew of N replication origin sequences, calculated using eq (\ref{cumsum}), from a single biological domain, are standardized to a single length, i.e., 16 or 8, by wavelet transformation, and are averaged to provide a single average skew numerical array. To identify whether an organism exhibits V-shaped pattern in RY- or MK-skew, we correlate the RY-skew detailed coefficient sequences of each organism with the average detailed coefficient sequence, calculated by averaging over the individual detailed coefficient sequences of all organisms within a single domain. We segregate an organism into RY- or MK-skew group depending on the correlation coefficient calculated above, and the general shape of the averaged approximate coefficients from RY-skew. If the latter is V-shaped and the correlation coefficient is negative, we assign the organism to MK-skew group, and vice versa. RY-skew $W_{RY}$ for the  RY-group organisms and MK-skew $W_{MK}$ for the MK-group organisms are calculated using eq (\ref{cumsum}). The final average skew is calculated by merging the two groups and averaging their skews. The dimensions of all the variables processed in the algorithm are mentioned below each of the steps.}}
    \label{fig: Algorithm}
\end{figure}

Solution for problem II: In order to solve this issue, we take our cue from the observation in genomic sequences that, for organisms that follow RY-grouping, the RY-cumulative skew exhibits a V-shaped curve, whereas MK-skew is usually an inverted-V or can be featureless as well (depending on the GC/AT ratio). Contrarily, for organisms following MK-grouping, MK-cumulative skew exhibits a V-shape and, RY-skew, an inverted-V or be featureless. This fact is also regularly used in origin-finding programs \cite{grigoriev1998analyzing, dong2023doric, sernova2008identification, zhang2005identification}. To group the organisms properly, after standardizing the lengths of replication origin sequences and taking wavelet transforms of RY-cumulative skew, we first take correlations of each of the last-level detailed coefficients, $D_i$, of all sequences (each denoted by `i') within a domain, with the averaged last-level detailed coefficient, $\langle D_i \rangle$. We create two groups of organisms (RY- and MK-skewed) within a domain, by segregating organisms that have positive correlation with the average, from organisms having negative correlation: i.e., $correlation(D_i,\langle D_i \rangle) < 0$ is clustered into a group, whereas, $correlation(D_i,\langle D_i \rangle) >= 0$ is clustered into another. To determine which of these two groups is RY-skewed, we look at the general shape of the average RY-cumulative skew of all the organisms in a domain. If the general shape is that of inverted-V, the average is dominated by MK-skew organisms, and hence, we assign all the positive correlation group to MK, and negative correlation group to RY. We switch the assignment in case of a V-shaped average cumulative skew. We then calculate the MK-skews for all the organisms assigned to the MK group, and RY-skews for the RY group, and take an average of both the skews to produce the final skew plot, for a given domain. This entire procedure is summarized in Fig. \ref{fig: Algorithm}.

\subsection*{Parameterization}

Our model involves only four parameters: $q$ and $r$, the formation and dissociation rates of hydrogen bonds between complementary bases, and $\alpha$ and $\beta$, the rate-modulating kinetic factors that are dependent on the orientation of the neighboring base pairs to the left and right. In the unzipping process, the formation and dissociation of base pairs are intramolecular and hence do not depend on base pair concentration \cite{dupuis2013single, bui2017design}. 
\begin{center}
    \ch{SS <=> [$K_{on}$][$K_{off}$]DS}
\end{center}
The ratio between the formation rate constant $k_{on}$ and dissociation rate constant $k_{off}$, the equilibrium constant $K_{eq}$, is purely thermodynamic and is dictated by the free energy difference between the bonded and unbonded states of base pairs through the equation \cite{hammes2012principles, hill1986introduction}   
\begin{equation}\label{eq constant}
    K_{eq} = exp({\frac{-\Delta{H}+T\Delta {S}}{RT}} ).
\end{equation}

The thermodynamic parameters $\Delta H$ and $\Delta S$ for GC base pairs (see below) are taken from Santalucia \textit{et al} as the average of all dimers of C and G, $\Delta H = \SI{-9.47}{Kcal\cdot mol^{-1}}$ and $\Delta S = \SI{-23.83}{cal \cdot mol^{-1} \cdot K^{-1}}$ \cite{santalucia2004thermodynamics} at PH 7 and $\SI{1}{M}$ NaCl. The corresponding melting point of the H-bond is $\SI{397.4}{K}$. Similarly, for AT base pairs, $\Delta H = \SI{-7.33}{Kcal\cdot mol^{-1}}$ and $\Delta S = \SI{-21.0}{cal \cdot mol^{-1} \cdot K^{-1}}$ and the corresponding melting point is $\SI{349.0}{K}$. We have chosen our parameters to be near phase transition, with temperature $\SI{360}{K}$for GC sequences and $\SI{312.4}{K}$ for AT sequences. Our results hold good near melting temperatures, and far below the melting point, they become length- and temperature-sensitive. For GC base pairs, the equilibrium rate constant $K_{eq}$, corresponding to temperature $\SI{360}{K}$ ($\approx T_m-\SI{37}{K}$), comfortably below melting temperature, is $3.48$ [\ref{eq constant}]. We take the rates of formation ($q$) and dissociation ($r$) to be $14 k_0$ and $4 k_0$ respectively, where $k_0$ is the scaling factor, of the order of $10^6 \, \text{s}$ \cite{altan2003bubble, manghi2016physics}. For AT base pairs, at a temperature $\SI{312.4}{K}$ ($\approx T_m-\SI{37}{K}$), these thermodynamic parameters ($K_{eq}$, $q$ and $r$) have nearly same values as that of GC base pairs. For creating the phase-space diagram, we vary the temperature, and therefore, $K_{eq}$ and $r$ are re-evaluated (keeping enthalpy-dominant $q$ constant) accordingly from eq (\ref{eq constant}). 

\par
As we are evaluating unzipping times of dsDNA, only sequence-dependent kinetic parameters are used in the work, since the sequence-independent asymmetric cooperativity stands canceled due to the anti-parallel orientations of the two DNA strands \cite{sequence_dependent}. The sequence-dependent asymmetric cooperativity catalytic and inhibiting factors used are $\alpha = 5$ and $\beta = 0.2$, respectively. It has to be strongly emphasized that the results below are insensitive to the precise values of the kinetic parameters and hold good for a wide range of the above parameters, with the constraint that $\alpha$ and $\beta$ values be unequal to instantiate kinetic asymmetry.  

\section*{Results}

We aim to compare the unzipping times of various sequences, fixing the sequence length and the unzipping temperature, for different lengths, and at various below-melting temperatures of dsDNA, and find the sequences that unzip the fastest, which we identify as putative replication origins. 

Since we are simulating the unzipping of linear DNA segments of a certain length, within a larger DNA double strand, the immediate neighborhood of the segment influences the unzipping rate. Depending on whether these boundary base pairs are kinetically catalyzing or inhibiting the base pairs at the two ends of the considered sequence, there are four possible boundary conditions. When both the boundary base pairs are catalyzing, the unzipping times of the sequences are faster compared to other boundary base pair orientations, and we use this boundary condition, as we are looking for the fastest unzipping sequences.

\begin{figure}[H]
\centering
\includegraphics[width=.6\textwidth]{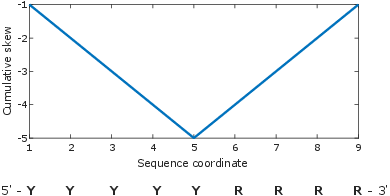}

    \caption{\textit{Cumulative skew plot of a 9-nt sequence $5'\text{-YYYYYRRRR-}3'$. The metric \enquote{cumulative skew} is computed as the difference between the number of purines and pyrimidines from the $5'$-end up to a certain location on the sequence, which is varied from the $5'$-to the $3'$-end, as expressed in eq (\ref{cumsum}). A highly-skewed RY-palindromic sequence shows a distinctive \enquote{V-shaped} cumulative skew plot as shown. Such V-shaped cumulative skews are seen in prokaryotic genomes, albeit at a much larger length scale (Mb), where they are regularly used to predict the location of the replication origin. }}
    \label{fig: Ideallyskewed}
\end{figure}

\begin{figure}[H]
\centering
\includegraphics[width=1\textwidth]{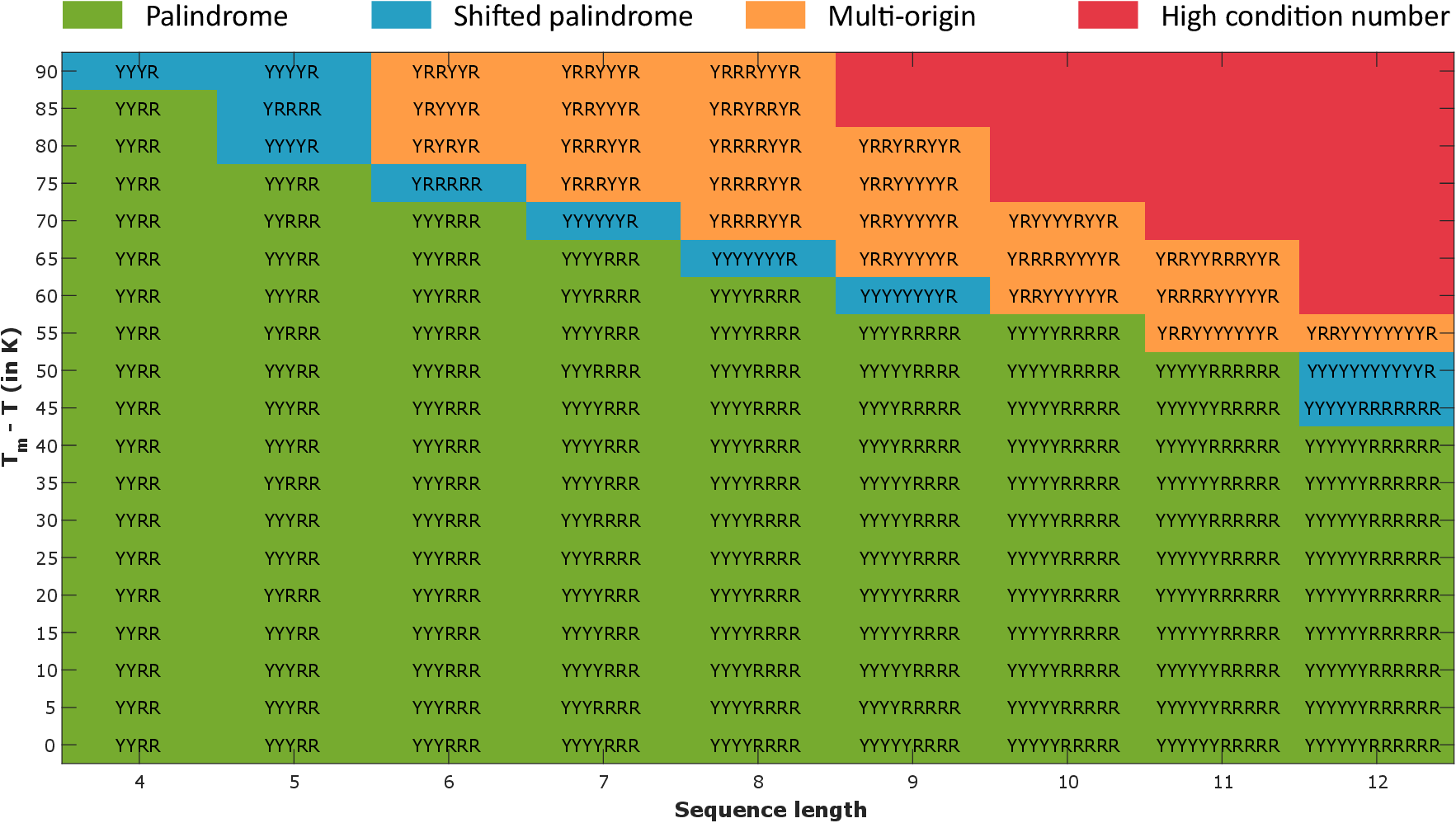}
    \caption{\textit{Evolutionarily superior, fast-unzipping sequences as a function of temperature and sequence length. At temperatures close to the melting temperature ($T_m$) of dsDNA, the fast-unzipping sequences of all lengths are entirely high-skew RY-palindromes, which allows for the parallel unzipping of both the palindromic arms. The regions in the `phase-space' where high-skew RY-palindromes emerge as winners are highlighted in green, which dominates the phase-space at high temperatures, for all lengths. At lower temperatures, we observe shifted-origin sequences, such as $5'\text{-YYRRRR-}3'$, dominating over all other sequences, particularly for short sequence lengths. This region, marked in blue, has however a much smaller phase-space area, and yields to multiple-origin sequences, marked in orange, upon further reduction in temperature. Multi-origin sequences can divide the sequences into more than two simultaneously-unzipping sections, thereby increasing the speed of replication at low temperatures, despite the presence of a high-barrier site between two origins. Further reduction in temperature or increase in length leads to rate matrices that are ill-conditioned, preventing us from identifying the nature of sequences at these locations of the phase-space. }}
    \label{fig: PhaseDiag}
\end{figure}

We find that near melting temperature of dsDNA, for a wide range of asymmetric cooperativity parameters around the ones chosen ($\alpha = 5$ and $\beta = 0.2$), of all the $64$ $6$-nucleotide-long sequences, the highly skewed (Fig. \ref{fig: Ideallyskewed}), RY-palindromic $5'\text{-YYYRRR-}3'$ sequence has the lowest unzipping time. This result holds for sequences of length up to 12 nt, with the sequence $5'\text{-YYYYYYRRRRRRR-}3'$ winning, beyond which inverting the large-dimensional rate matrix becomes computationally intractable. On lowering the temperature well below the melting temperature, multi-origin sequences such as $5'\text{-YYRRYYRR-}3'$ and $5'\text{-YYYRRYYR-}3'$ become the fastest unzipping sequences. The dependence of the winning sequence on the ambient temperature and sequence length is shown in the `phase-space' diagram, Fig. \ref{fig: PhaseDiag}.

\subsection*{Unzipping begins at low kinetic barrier locations}
\begin{figure}[H]
\centering
\includegraphics[width=\textwidth]{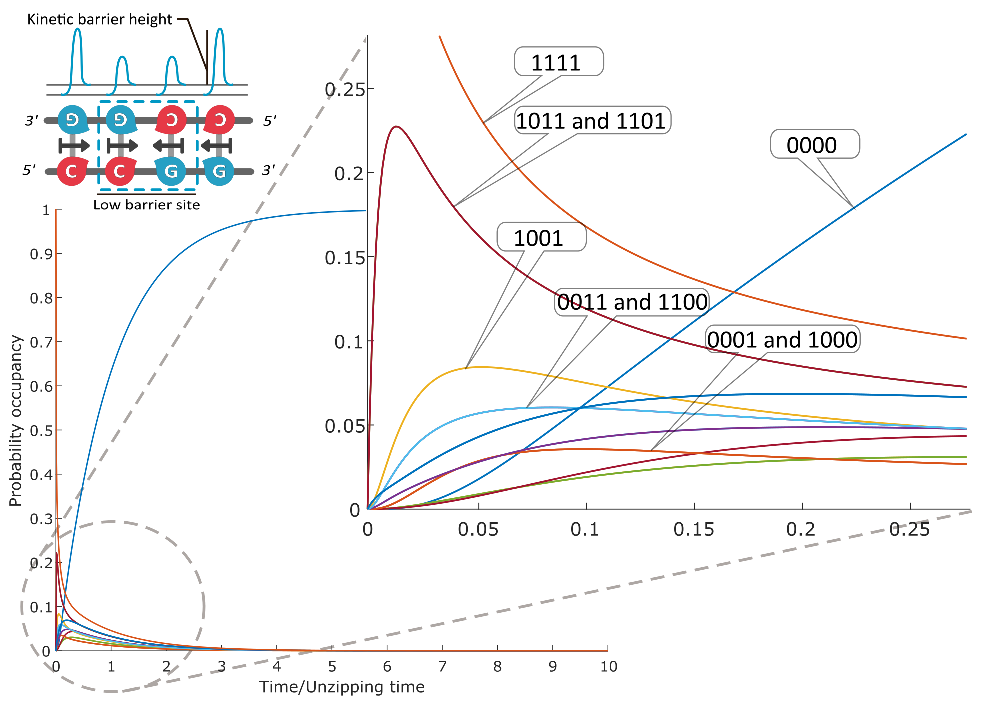}

    \caption{\textit{ Time evolution of probabilities of occupation of different states in the Markov chain constructed to evaluate the unzipping times of the sequence $5'\text{-CCGG-}3'$. The probabilities are evaluated for the parameters $q=14k_0$, $ r=10k_0$, $\alpha=5$ and $\beta=0.2$. This sequence has a low kinetic barrier site at the second and third H-bond locations. The Markov chain is initially in the fully zipped $1111$ state, and therefore, its occupation probability is 1 at time $t=0$. As time increases, the occupation probabilities of various states peak at various times, thereby providing insight into the unzipping process itself. The states that peak immediately after the beginning of unzipping are $1011$ and $1101$, demonstrating that the second and the third H-bonds are the first to break. The next most probable state visited by the Markov chain is $1001$ which shows the unzipping bubble expanding towards the left and right, due to the dynamically altered low barrier heights of the bonds at the right and left edges of the bubble. The states $0011$ and $1100$ peak next. Eventually, the system reaches the fully unzipped state ($0000$) with the probability of 1 at the steady state. This figure illustrates the unzipping process of a skewed RY-palindrome as a sequential, bidirectional breaking of H-bonds, beginning at the center, and thus reproduces the behavior of origins of replication. }}
    \label{fig: time evolution of probability}
\end{figure}

\begin{figure}[H]
     \centering
     \includegraphics[width = 0.6\textwidth]{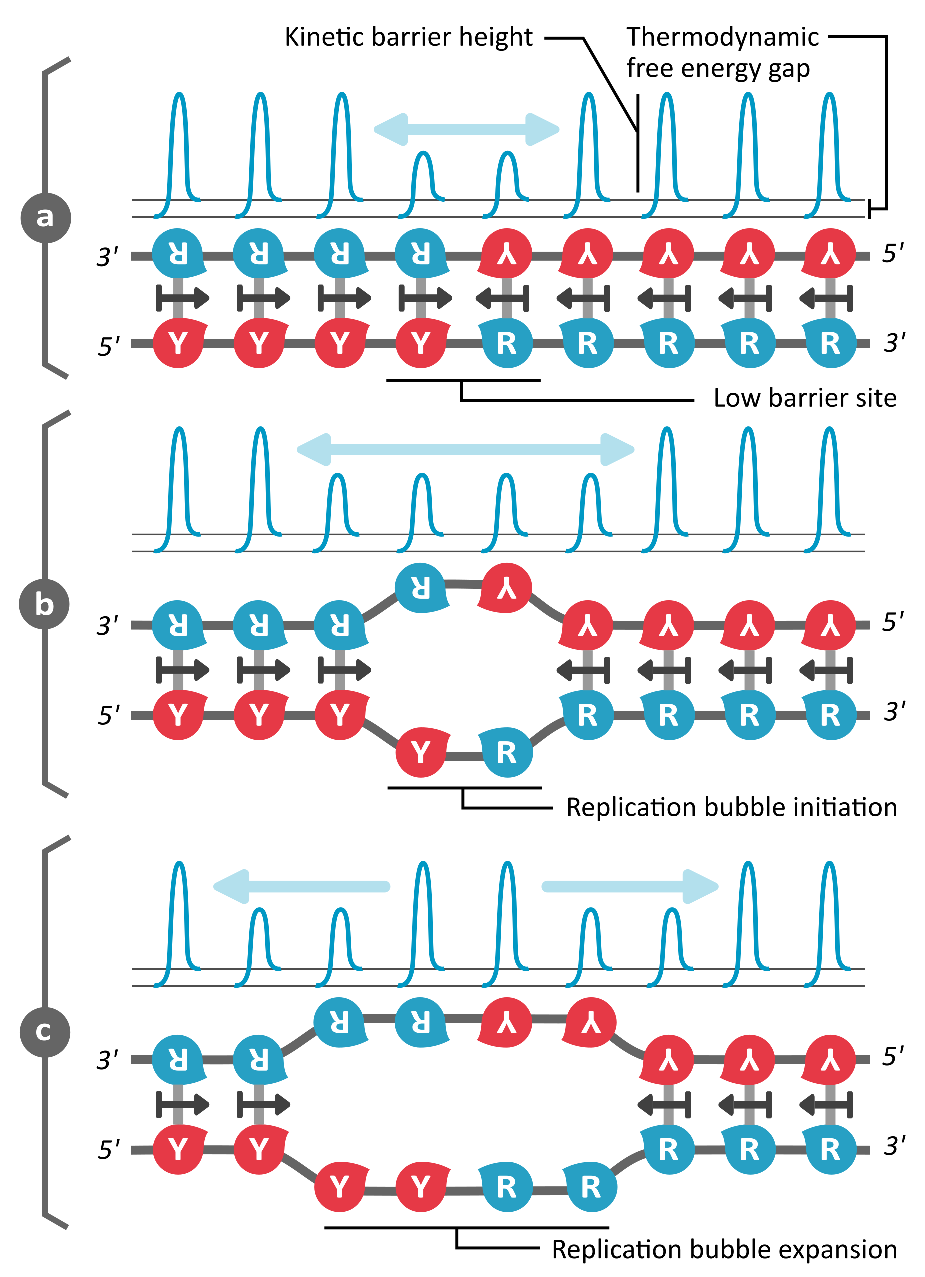}
         
     \caption{\textit{ Schematic illustration of the unzipping process of the high-skew, RY-palindromic sequence $5'\text{-YYYYRRRRR-}3'$, functioning as a replication origin site. (a) The kinetic barriers for the formation/dissociation of all nine hydrogen bonds are shown above the double strands. These barriers are sequence-dependent, with the barriers of the fourth and fifth bonds from the left (highlighted with a blue arrow) being weaker, due to the mutually catalytic asymmetric cooperativities, as shown by the directions of the black arrows at these two locations, drawn between the two strands. (b) These low kinetic barrier hydrogen bond locations are prone to break, resulting in the formation of a single-stranded bubble. Due to the absence of the inhibitory influence of the recently-broken fourth and fifth hydrogen bonds on the third and sixth hydrogen bonds, the kinetic barriers of these later bonds are dynamically reduced, leaving these two bonds weaker than the rest of the bonds. (c) The breaking of hydrogen bonds progresses towards either end of the sequence, with the low kinetic barrier third and sixth bonds dissociating, which in turn dynamically reduces kinetic barriers of second and seventh hydrogen bonds from the left. The panels (a), (b) and (c) thus illustrates the sequence of bond-breaking events guided by the dynamic alteration of kinetic barriers, and shows the bidirectional unzipping of the RY-palindromic sequence, capturing the observed prokaryotic replication origin behavior. It should be clear from the figure that high nucleotide skews are essential for this sequence-dependent local unzipping. }}  
     \label{fig: Bubble formation}
\end{figure}

In order to understand the above results, we evaluate the time-evolution of the probabilities of the Markov chain to be in a specific state, for all states, using the equation (\ref{evolution of probability}). We choose the sequence $5'\text{-CCGG-}3'$, to be concrete. 
At time $t=0$, the probability of the fully-zipped state $1111$ is 1, since we are starting at the fully-zipped state, as seen in Fig. \ref{fig: time evolution of probability}. As time progresses, the first hydrogen bond to break is either the second or third, i.e., $1101$ or $1011$, since the kinetic barriers of these two bonds are the lowest of all, as seen in the barrier diagram in Fig. \ref{fig: time evolution of probability}. As the probability of the $1111$ state decreases, the probability of $1101$ and $1011$ states increases, and the barrier heights of the hydrogen bonds are dynamically altered due to the absence of the second or the third bond, as illustrated in Fig. \ref{fig: Bubble formation}. The barrier heights of the second and the fourth hydrogen bonds in $1101$ or the first and third bonds in $1011$ are now the lowest, and the probability of occupation of these states, i.e., $1001$, $1100$, and $0011$, peak next. The next bonds to break are the ones that are adjacent to the two already-broken H-bonds, thereby expanding the replication bubble. One can gather that the unzipping process begins at the sequence-determined low barrier location in the middle of the sequence, and proceeds in both directions simultaneously until the entire sequence is unzipped. Therefore, the low barrier location at the center of the sequence $5'\text{-CCGG-}3'$ functions as an \enquote{origin of replication}. This observation also quantitatively corroborates our qualitative arguments in \cite {sequence_dependent}, where we have argued that anti-parallel DNA strands are evolutionarily beneficial because they parallelize the self-replication process by dividing the genome into two halves (called replichores) which replicate independently and simultaneously. The above observations obviously hold good for longer sequences (e.g. $5'\text{-YYYRRR-}3'$, $5'\text{-YYYRRRR-}3'$) and sequences with shifted origin sites (e.g. $5'\text{-YYRRRRR-}3'$, $5'\text{-YYYYYRR-}3'$, $5'\text{-YYRRRR-}3'$).

As a corollary to the above observation that unzipping begins at low barrier locations of the sequence, we can show that the last regions to unzip are the high barrier locations. We can demonstrate this by flipping the direction of the $5'\text{-CCGG-}3'$ sequence to $5'\text{-GGCC-}3'$, which now has a high barrier location in the middle of the sequence, due to the mutually stabilizing influence of sequence-dependent asymmetric cooperativity of the middle $GC$ base-pair. A spectral decomposition (eq \ref{spectral terms}) of the rate matrix decomposes the unzipping process into \enquote{modes} of unzipping with different rates of unzipping for different locations. We observe that the eigenvectors corresponding to the lowest eigenvalues (slowest rates) have significant components at the high barrier locations, indicating that these locations unzip at the slowest rate. We double-checked this by evaluating the time evolution of probabilities, where the probabilities of states with high barrier locations peak much later than other states. Since the contribution of high barrier locations to the unzipping time of a given sequence is significantly large, we conclude that sequences functioning as origins of replication discourage the presence of high barrier sub-sequences. In addition to the above, we observe that the presence of a stretch of homogeneous sequence, composed entirely of purines or pyrimidines, exhibiting asymmetric cooperativity in the same direction, (e.g. $5'\text{-YYYY-}3'$, or $5'\text{-RRRR-}3'$, etc.)  facilitates unidirectional sequential unzipping and elongation of the replication bubble, thereby expediting the unzipping and replication process. Therefore, such sequences are seen on either side of the low-barrier sites in our results. The temporal progression of unzipping of a high-skew RY-palindromic sequence is schematically illustrated in Fig. \ref{fig: Bubble formation}.

\par

\begin{figure} [H]
     \centering
     \hspace*{-0.15\textwidth}
     \includegraphics[width = 1.3\textwidth, trim=50pt 50pt 60pt 20pt, clip]{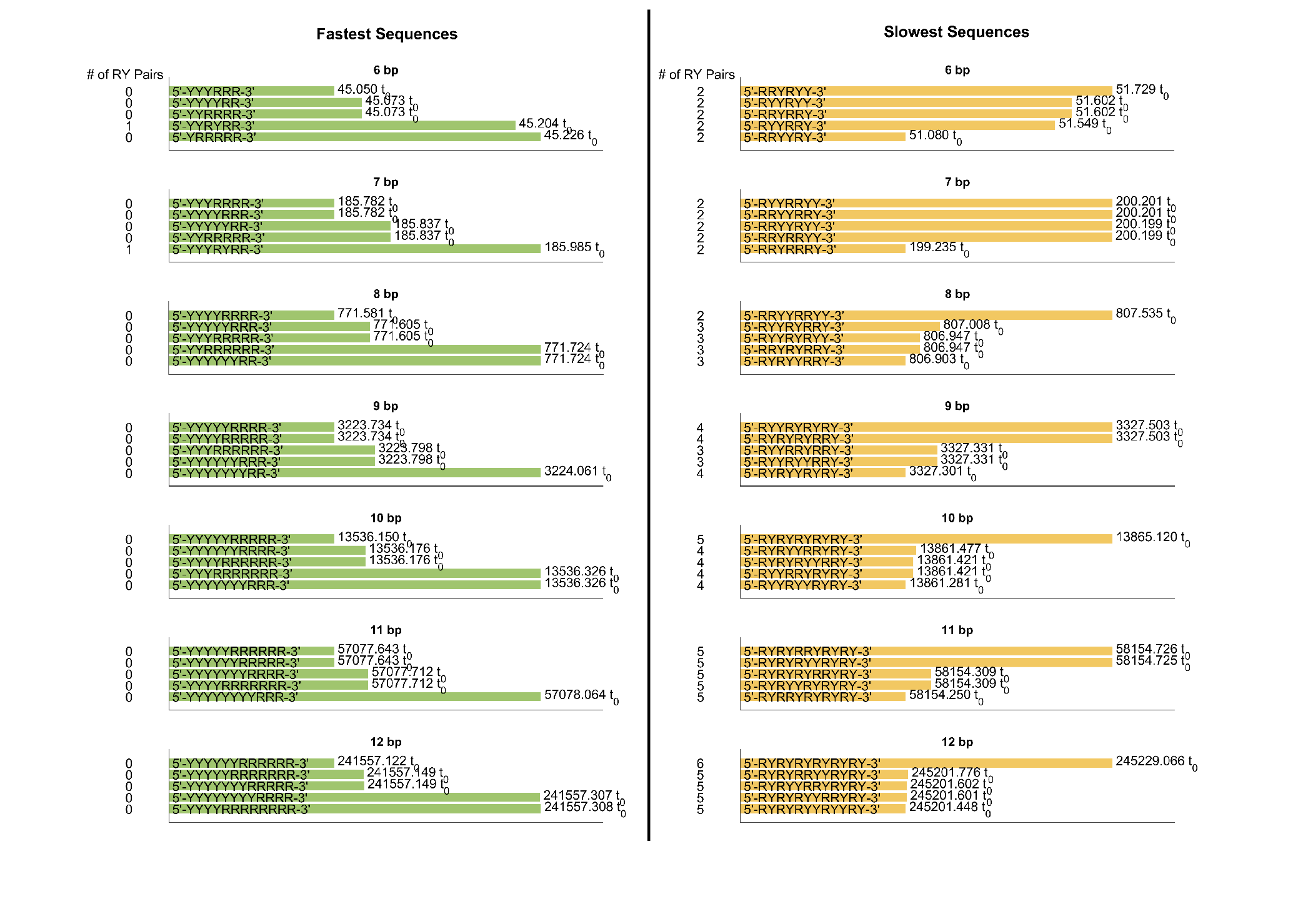}
     
     \caption{\textit{ Variation in characteristics of the evolutionarily superior and inferior sequences with variation in sequence length, at a constant temperature of $T_m -\SI{37}{K}$ (not to scale). At $T_m -\SI{37}{K}$, the rate matrices for both GC and AT sequences are the same, and therefore the sequences are generalized as RY sequences. The time taken for unzipping of the top five fastest (left panel) and slowest (right panel) sequences for various sequence lengths is provided, in units of $t_0 = 1/k_0$, where $k_0$ is the rate scaling factor, of the order of $10^6 \, \text{s}$. Note the high skews of the fastest-unzipping sequences and the lack of skews in the slowest sequences, skew being defined as an excess of purines over pyrimidines over one-half of the sequence. The fastest sequences of all lengths are RY-palindromic. The rapid unzipping of high-skew, RY-palindromic sequences is due to parallelization of the unzipping process by dividing the sequence into two simultaneously unzipping sections, beginning at the central low-barrier site, as shown in Fig. \ref{fig: Bubble formation}. A central characteristic of the fastest (slowest) sequences is the absence (presence) of the mutually-inhibiting, high-kinetic-barrier dinucleotide sequence $5'\text{-RY-}3'$, and the presence (absence) of the mutually-catalytic, low-barrier dinucleotide sequence $5'\text{-YR-}3'$. The total number of the high-barrier $5'\text{-RY-}3'$ dinucleotides in each sequence is provided to the left of the sequences to illustrate the deleterious effect of high-barrier sites on the unzipping time. These arguments apply to MK-palindromes as well, if the asymmetric cooperativity direction of AT base pair is switched.}}
     \label{fig: fastSeqs}
\end{figure}

The evolutionary competition between various sequences, of same length and at same temperature, for resources such as monomer supply and energy sources, results in some sequences dominating the fitness landscape, due to their ability to replicate faster. Since unzipping of dsDNA is the rate-limiting step for replication at below-melting temperatures, sequences that unzip faster than others hold an evolutionary advantage at these temperatures. Above, we have argued that, near the melting temperatures of dsDNA, high-skew RY- (or MK-)palindromic sequences unzip faster and hence emerge as winners of the evolutionary competition. It is instructive to visualize the fastest- and slowest-unzipping sequences, as a function of sequence length and temperature, in order to gain insight into the variations in the characteristics of the fastest sequences as the temperature is lowered or the sequence length is increased. The list of the top five fastest- and slowest-unzipping sequences at a constant temperature of $T_m -\SI{37}{K}$ (applicable for both GC and AT sequences), when the sequence length is varied from 6 nt to 12 nt, is shown in Fig. \ref{fig: fastSeqs}. At these near-melting temperatures, the fastest-unzipping sequences are either high-skew RY-palindromes, such as $5'\text{-YYYYRRRR-}3'$, or high-skew near-palindromes with shifted origins, such as $5'\text{-YYYRRRRR-}3'$ and $5' \text{-YYRRRRRR-} 3'$.

However, when the temperature is substantially lowered below the melting temperatures of dsDNA, multi-origin sequences (e.g., $5'\text{-YYRRYYRR-}3'$) begin to unzip faster than the single-origin ones. As the temperature is lowered, unzipping becomes more unfavorable, and the introduction of multiple origins helps bring down unzipping time, even though it simultaneously introduces a single high barrier location between every pair of origins. At low temperatures, the ability of low-barrier origins in bringing down the unzipping time is significantly more than the ability of high-barrier locations in increasing the unzipping time. The converse is true at temperatures close to the melting temperatures: high-barrier locations increase the unzipping time significantly more than low-barrier regions are capable of bringing it down, and hence are avoided in the fastest sequences at such temperatures. It has to be remembered that these conclusions are valid when only the nearest-neighbor asymmetric cooperativity interactions are included. When these cooperativity interactions are extended to next-nearest-neighbors and beyond, we expect the phase space of single-origin sequence winners to expand towards larger lengths and lower temperatures. Furthermore, we observe that the unzipping time increases exponentially with increasing sequence length, as has been experimentally observed in dsDNA denaturation and unzipping experiments \cite{dupuis2013single, porschke1971co, woodside2006nanomechanical}.

As mentioned in the introduction, we have avoided sequences containing a mixture of all four nucleotides, and restricted ourselves to sequences containing either purely GC or purely AT base pairs. The reason is that, in a direct competition between GC-only sequences and AT-only sequences, the AT sequences will win, simply because of lower thermodynamic barrier. However, experimental data below show that most replication origin sequences have a nearly balanced GC and AT distribution, with a moderate preference for AT (about 20\% more than GC in bacteria). If replication rate was the only selective factor, then these sequences would be highly AT-rich. This discrepancy between the model results and the experimentally observed data is due to non-inclusion of the need for information storage in DNA sequences in the model. Replication rate and information storage potential, when considered together, leads to a balanced distribution of GC and AT base pairs in the origin sequences, seen experimentally, and as we show in an upcoming article. This need for information storage also reduces the magnitude of skew in eukaryotes, since their information storage requirements are comparatively larger. Therefore, by including only replication rate as the selective factor and ignoring information storage, we have made our model quite restrictive, resulting in its inability to address the evolutionary superiority of mixed nucleotide, high-skew, palindromic sequences.

\section*{Experimental Support}

Bioinformatic identification of origins of replication in both Prokaryotes and Eukaryotes typically involves finding genomic locations where cumulative GC/RY/MK skew has a prominent minimum \cite{grigoriev1998analyzing, frank2000oriloc, lobry1996asymmetric, luo2019doric, gao2008ori}. The cumulative skew of a sequence is a sequence of integers of the same length as that of the genomic sequence, defined as a running cumulative sum of the difference between the number of G/R/K's and the number of C/Y/M's, expressed mathematically in eq (\ref{cumsum}) for RY\cite{zhang2003z, neccsulea2007new}. 

This approach to finding the origins of replication implies an asymmetric distribution of G/R/K's and C/Y/M's around the origin, with more C/Y/M's towards the $5'$-end and more G/R/K's towards the $3'-$end. This is precisely what we find as the signature of the fastest unzipping sequences within our model, although at a much shorter length scale. Near melting temperatures, the winning sequences in our model exhibit maximal skews, with the $5'-$end entirely made up of C/Y/M's and the $3'$-end, of G/R/K's. The location where the skew switches from C/Y/M-dominant to G/R/K-dominant, where the cumulative skew exhibits a minimum, is the location of the origin of replication, as we showed above in the `Results' section, and illustrated in the Fig. \ref{fig: Ideallyskewed}. \textit{This is precisely what is seen in the experimentally determined replication origins of mitochondrial light strand origins}, as we show below. Thus, our model provides both mechanistic and evolutionary rationale for the existence of skews around replication origins. The choice between RY- and MK-skew is dictated by the asymmetric cooperativity direction of AT base pair in an individual organism, which varies across organisms.

\par
In order to examine if significant skews are present around replication origins, we downloaded $5686$ mitochondrial replication origin sequences from the NCBI database (https://www.ncbi.nlm.nih.gov/nuccore), and $227532$ bacterial, $851$ plasmid and $801$ archaeal replication origin sequences from DoriC database (https://tubic.org/doric/). The average lengths of these OriC sequences are 30 nt, 413 nt, 644 nt and 576 nt for mitochondria, bacteria, archaea and plasmid, respectively. We employ a wavelet-based method to reduce the length of sequences to a uniform length of 16 (8 for mitochondria), as described in the Methods section. Our method allows us to concentrate on the large-scale structures present in long origin sequences present in the above databases, by removing information at smaller length scales. This procedure helps find signatures common to both mitochondrial origins of length scale of tens of base pairs, and bacteria, of length scales of the order of hundreds to thousands of base pairs. We also account for different modes of asymmetric cooperativity of AT base pairs in different organisms, by segregating them into two groups (RY and MK). We find, \textit{length-scale-invariant}, near-identical, V-shaped signature of replication origin in all the four domains / cellular components studied, in complete consonance with our theoretical model.

\begin{figure}[H]
    \centering
        \centering
        \includegraphics[width=\textwidth]{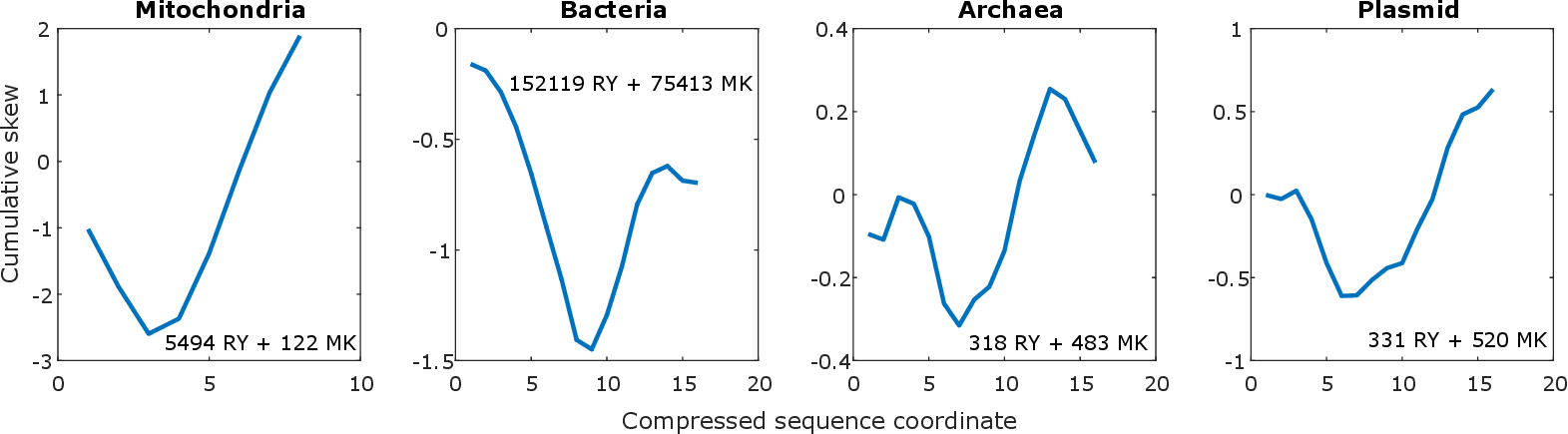}
    \caption{\textit{Average cumulative skew of (a) 5616 mitochondria (from NCBI database) (b) 227532 bacteria (c) 801 archaea and (d) 851 plasmid replication origin sequences (from DoriC database) are shown here. The sequences with length ranging from $32 nt$ ($16 nt$ in case of mitochondria) to $1000 nt$ are extracted and processed according to the procedure described in the Methods section. In case of mitochondria, to the left of the origin, denoted by the tip of the V-shaped curve, 83\% (5 out of 6) of nucleotides are pyrimidine/amino (Y/M) nucleotides, and to the right, 85\% (8.5 out of 10) are purine/keto (R/K) nucleotides. In case of bacteria, this asymmetry is weaker in comparison to mitochondria. Near the minima, to the left of the origin, there is a 11\% excess Y/M, and to the right, 8\% excess R/K nucleotides. These asymmetries in RY or MK content result in a low-barrier site at the origin, and lead to its early dissociation and subsequent parallel unzipping of the two arms of the sequence, thereby instantiating replication origin. For archaea, the enrichment percentages to the right and left are 7.5\% and 10\%, respectively. For plasmids, these percentages are 12\% (left) and 14\% (right).}}
    \label{fig: Evidence}
\end{figure}

The average cumulative skew of mitochondria, bacteria, archaea and plasmid OriC sequences are shown in Fig. \ref{fig: Evidence}. In mitochondria, almost all the sequences (5494 out of 5686) belong to RY grouping, and the skew is comparatively stronger than that of bacteria, archaea and plasmid. In mitochondria, around $83\%$ (5 out of 6) bases to the $5'$-end of the origin are found to be pyrimidines, and to the $3'$-end around 85\% (8.5 out of 10) of the bases are purines. In bacteria, near the skew minima, we find around 11\% excess Y/M to the $5'$-end and 8\% excess R/K to the $3'$-end. In the case of archaea and plasmids, around 60\%  OriC sequences belong to MK grouping and the rest belong to RY grouping. Their enrichment percentages are 7.5\% and 12\% near the $5'$-end of the origin and 10\% and 14\% to the $3'$-end of the origin, respectively. 

\par
It is generally argued that palindromes exert their functionalities, including that of origin, through the formation of stem-loop structures \cite{bartholdy2015allele, cayrou2011genome, cayrou2012new, bikard2010folded, pearson1996inverted, cheung2004palindrome}. Our explanation for the origin functionality contradicts this argument and suggests that origins are determined by low kinetic barriers for hydrogen bond dissociation, determined by sequence-dependent asymmetric cooperativity. An elegant experiment \cite{wanrooij2012vivo} by S Wanrooij \textit{et al} has demonstrated, in the highly-skewed mitochondrial light strand origin sequences, that the origin functionality can be entirely abrogated by switching the two arms of the palindromic sequence (construct (e) in the paper), leaving the $5'$-arm purine-abundant and $3'$-arm pyrimidine-abundant. Although the resultant sequence still was capable of forming a stem-loop structure, there was no evidence of origin functionality of the sequence, thus demonstrating that the ability to form stem-loop structures is inconsequential to origin functionality. Instead, the direction of skew change, from $5'$-pyrimidine to $3'$-purine, and the consequent presence of $5'\text{-YR-}3'$, results in a low barrier site and lead to origin functionality. Whereas, when the skew is flipped across the origin, with the skew changing from $5'$-purine to $3'$-pyrimidine, and the consequent presence of $5'\text{-RY-}3'$, a high barrier site results, which abrogates origin functionality.

\section*{Discussion}
In our earlier paper \cite{sequence_dependent}, we hypothesized, using qualitative arguments, that DNA \textit{parallelizes} replication by loading the leading strand of each replichore (section of genome between the replication origin and the terminus) with pyrimidines, thereby dictating the direction that the unzipping machinery should move in, through sequence-dependent asymmetric cooperativity. This leads to the two replichores of the prokaryotic genome, skewed in nucleotide content, replicating independently and simultaneously, which increases the rate of replication, making such sequences evolutionarily superior. In this study, we \textit{quantitatively} show that sequences that are loaded with purine/keto nucleotides on $3'$-half and pyrimidine/amino nucleotides on the $5'$-half, i.e., sequences with large V-shaped RY- or MK-skews, such as $5'\text{-YYYYRRRR-}3'$, replicate faster than other sequences of the same length, due to their faster unzipping ability. The choice of RY or MK skew is dictated by the direction of asymmetric cooperativity of AT base pair, which varies across organisms, even within the same domain. We have shown that the faster replication is due to the sequence-determined low kinetic barriers at locations such as $5'\text{-YR-}3'$ (or $5'\text{-MK-}3'$), as illustrated in Fig. \ref{fig: Bubble formation}, that function as replication origins. At temperatures much lower than the melting temperatures of dsDNA sequences, the fast-unzipping sequences are not high-skew RY-palindromes, but sequences with more number of low-barrier sites, with high-barrier sites necessarily interspersed among them. We have evaluated the unzipping times of various sequences using the \enquote{Continuous Time Markov Chain} method, after appropriately parameterizing the thermodynamic and kinetic variables of the system. Crucially, we have used just two free parameters in our model. Our results are not sensitive to the parameters chosen, and hence can be generalized. As experimental evidence, we have analyzed 5686 mitochondrial, 227532 bacterial, 801 archaeal and 851 plasmid replication origin sequences, where we have found a significant asymmetry in purine-pyrimidine/amino-keto composition across $3'$ and $5'$ ends, just as predicted by our model. We have also pointed to an experiment where switching the purine-pyrimidine asymmetry across the origin abrogates the origin functionality, again in complete consonance with our model. 

More generally, our computational experiments employing asymmetric cooperativity model have demonstrated the importance of DNA sequence in determining its evolutionary superiority, by influencing the sequence's unzipping kinetics (and most probably its replication kinetics as well). One can imagine, in the prebiotic scenario, of evolutionary competition among various self-replicating molecules for monomers and energetic sources, where, self-replicators that replicated faster won the evolutionary race. By connecting the self-replicators' rate of replication with their sequence, our model allows for the emergence of biological information. This study envisages an important role for high-skew RY-/MK-palindromic sequences in the prebiotic scenario, where such sequences could both replicate faster and, due to their internal sequence complementarity, could form hairpin loops to carry out catalytic functions. Since the functioning of transposons (or mobile genetic elements) and CRISPR arrays depend on local unzipping, we anticipate that our model applies to them as well, and a similar analysis for such sequences is warranted.


\printbibliography

@article{lee2023and,
  title={Where and when to start: Regulating DNA replication origin activity in eukaryotic genomes},
  author={Lee, Clare SK and Wei$\beta$, Matthias and Hamperl, Stephan},
  journal={Nucleus},
  volume={14},
  number={1},
  pages={2229642},
  year={2023},
  publisher={Taylor \& Francis}
}

@article{prioleau2016dna,
  title={DNA replication origins—where do we begin?},
  author={Prioleau, Marie-No{\"e}lle and MacAlpine, David M},
  journal={Genes \& development},
  volume={30},
  number={15},
  pages={1683--1697},
  year={2016},
  publisher={Cold Spring Harbor Lab}
}

@article{cayrou2011genome,
  title={Genome-scale analysis of metazoan replication origins reveals their organization in specific but flexible sites defined by conserved features},
  author={Cayrou, Christelle and Coulombe, Philippe and Vigneron, Alice and Stanojcic, Slavica and Ganier, Olivier and Peiffer, Isabelle and Rivals, Eric and Puy, Aurore and Laurent-Chabalier, Sabine and Desprat, Romain and others},
  journal={Genome research},
  volume={21},
  number={9},
  pages={1438--1449},
  year={2011},
  publisher={Cold Spring Harbor Lab}
}

@article{cayrou2012new,
  title={New insights into replication origin characteristics in metazoans},
  author={Cayrou, Christelle and Coulombe, Philippe and Puy, Aurore and Rialle, Stephanie and Kaplan, Noam and Segal, Eran and M{\'e}chali, Marcel},
  journal={Cell cycle},
  volume={11},
  number={4},
  pages={658--667},
  year={2012},
  publisher={Taylor \& Francis}
}

@article{bartholdy2015allele,
  title={Allele-specific analysis of DNA replication origins in mammalian cells},
  author={Bartholdy, Boris and Mukhopadhyay, Rituparna and Lajugie, Julien and Aladjem, Mirit I and Bouhassira, Eric E},
  journal={Nature communications},
  volume={6},
  number={1},
  pages={7051},
  year={2015},
  publisher={Nature Publishing Group UK London}
}

@article{mechali2001dna,
  title={DNA replication origins: from sequence specificity to epigenetics},
  author={M{\'e}chali, Marcel},
  journal={Nature reviews genetics},
  volume={2},
  number={8},
  pages={640--645},
  year={2001},
  publisher={Nature Publishing Group UK London}
}

@article{lobry1996asymmetric,
  title={Asymmetric substitution patterns in the two DNA strands of bacteria.},
  author={Lobry, Jean R},
  journal={Molecular biology and evolution},
  volume={13},
  number={5},
  pages={660--665},
  year={1996}
}

@article{porschke1971co,
  title={Co-operative non-enzymatic base recognition III. Kinetics of the helix—coil transition of the oligoribouridylic{\textperiodcentered} oligoriboadenylic acid system and of oligoriboadenylic acid alone at acidic pH},
  author={P{\"o}rschke, Dietmar and Eigen, Manfred},
  journal={Journal of molecular biology},
  volume={62},
  number={2},
  pages={361--381},
  year={1971},
  publisher={Elsevier}
}

@article{bui2017design,
  title={Design and analysis of localized DNA hybridization chain reactions},
  author={Bui, Hieu and Miao, Vincent and Garg, Sudhanshu and Mokhtar, Reem and Song, Tianqi and Reif, John},
  journal={Small},
  volume={13},
  number={12},
  pages={1602983},
  year={2017},
  publisher={Wiley Online Library}
}

@article{woodside2006nanomechanical,
  title={Nanomechanical measurements of the sequence-dependent folding landscapes of single nucleic acid hairpins},
  author={Woodside, Michael T and Behnke-Parks, William M and Larizadeh, Kevan and Travers, Kevin and Herschlag, Daniel and Block, Steven M},
  journal={Proceedings of the National Academy of Sciences},
  volume={103},
  number={16},
  pages={6190--6195},
  year={2006},
  publisher={National Acad Sciences}
}

@book{gillespie1991markov,
  title={Markov processes: an introduction for physical scientists},
  author={Gillespie, Daniel T},
  year={1991},
  publisher={Elsevier}
}

@article{frank2000oriloc,
  title={Oriloc: prediction of replication boundaries in unannotated bacterial chromosomes},
  author={Frank, AC and Lobry, JR},
  journal={Bioinformatics},
  volume={16},
  number={6},
  pages={560--561},
  year={2000},
  publisher={Oxford University Press}
}

@article{luo2019doric,
  title={DoriC 10.0: an updated database of replication origins in prokaryotic genomes including chromosomes and plasmids},
  author={Luo, Hao and Gao, Feng},
  journal={Nucleic acids research},
  volume={47},
  number={D1},
  pages={D74--D77},
  year={2019},
  publisher={Oxford University Press}
}

@article{gao2008ori,
  title={Ori-Finder: a web-based system for finding oriC s in unannotated bacterial genomes},
  author={Gao, Feng and Zhang, Chun-Ting},
  journal={BMC bioinformatics},
  volume={9},
  number={1},
  pages={1--6},
  year={2008},
  publisher={BioMed Central}
}

@article{pabo1984protein,
  title={Protein-DNA recognition},
  author={Pabo, Carl O and Sauer, Robert T},
  journal={Annual review of biochemistry},
  volume={53},
  number={1},
  pages={293--321},
  year={1984},
  publisher={Annual Reviews 4139 El Camino Way, PO Box 10139, Palo Alto, CA 94303-0139, USA}
}

@article{chua2012mechanics,
  title={The mechanics behind DNA sequence-dependent properties of the nucleosome},
  author={Chua, Eugene YD and Vasudevan, Dileep and Davey, Gabriela E and Wu, Bin and Davey, Curt A},
  journal={Nucleic acids research},
  volume={40},
  number={13},
  pages={6338--6352},
  year={2012},
  publisher={Oxford University Press}
}

@article{kells2020correlation,
  title={Correlation functions, mean first passage times, and the Kemeny constant},
  author={Kells, Adam and Koskin, Vladimir and Rosta, Edina and Annibale, Alessia},
  journal={The Journal of Chemical Physics},
  volume={152},
  number={10},
  year={2020},
  publisher={AIP Publishing}
}

@article{dupuis2013single,
  title={Single-molecule kinetics reveal cation-promoted DNA duplex formation through ordering of single-stranded helices},
  author={Dupuis, Nicholas F and Holmstrom, Erik D and Nesbitt, David J},
  journal={Biophysical journal},
  volume={105},
  number={3},
  pages={756--766},
  year={2013},
  publisher={Elsevier}
}

@article{wanrooij2012vivo,
  title={In vivo mutagenesis reveals that OriL is essential for mitochondrial DNA replication},
  author={Wanrooij, Sjoerd and Miralles Fust{\'e}, Javier and Stewart, James B and Wanrooij, Paulina H and Samuelsson, Tore and Larsson, Nils-G{\"o}ran and Gustafsson, Claes M and Falkenberg, Maria},
  journal={EMBO reports},
  volume={13},
  number={12},
  pages={1130--1137},
  year={2012},
  publisher={John Wiley \& Sons, Ltd Chichester, UK}
}

@article{bell1992atp,
  title={ATP-dependent recognition of eukaryotic origins of DNA replication by a multiprotein complex},
  author={Bell, Stephen P and Stillman, Bruce},
  journal={Nature},
  volume={357},
  number={6374},
  pages={128--134},
  year={1992},
  publisher={Nature Publishing Group UK London}
}

@article{lee1997architecture,
  title={Architecture of the yeast origin recognition complex bound to origins of DNA replication},
  author={Lee, Daniel G and Bell, Stephen P},
  journal={Molecular and cellular biology},
  year={1997},
  publisher={Am Soc Microbiol}
}

@article{fuller1983purified,
  title={Purified dnaA protein in initiation of replication at the Escherichia coli chromosomal origin of replication.},
  author={Fuller, Robert S and Kornberg, Arthur},
  journal={Proceedings of the National Academy of Sciences},
  volume={80},
  number={19},
  pages={5817--5821},
  year={1983},
  publisher={National Acad Sciences}
}

@article{grigoriev1998analyzing,
  title={Analyzing genomes with cumulative skew diagrams},
  author={Grigoriev, Andrei},
  journal={Nucleic acids research},
  volume={26},
  number={10},
  pages={2286--2290},
  year={1998},
  publisher={Oxford University Press}
}

@article{perna1995patterns,
  title={Patterns of nucleotide composition at fourfold degenerate sites of animal mitochondrial genomes},
  author={Perna, Nicole T and Kocher, Thomas D},
  journal={Journal of molecular evolution},
  volume={41},
  pages={353--358},
  year={1995},
  publisher={Springer}
}

@article{charneski2011atypical,
  title={Atypical AT skew in Firmicute genomes results from selection and not from mutation},
  author={Charneski, Catherine A and Honti, Frank and Bryant, Josephine M and Hurst, Laurence D and Feil, Edward J},
  journal={PLoS Genetics},
  volume={7},
  number={9},
  pages={e1002283},
  year={2011},
  publisher={Public Library of Science San Francisco, USA}
}

@book{hammes2012principles,
  title={Principles of chemical kinetics},
  author={Hammes, Gorden},
  year={2012},
  publisher={Elsevier}
}

@article{altan2003bubble,
  title={Bubble dynamics in double-stranded DNA},
  author={Altan-Bonnet, Gr{\'e}goire and Libchaber, Albert and Krichevsky, Oleg},
  journal={Physical review letters},
  volume={90},
  number={13},
  pages={138101},
  year={2003},
  publisher={APS}
}

@article{manghi2016physics,
  title={Physics of base-pairing dynamics in DNA},
  author={Manghi, Manoel and Destainville, Nicolas},
  journal={Physics Reports},
  volume={631},
  pages={1--41},
  year={2016},
  publisher={Elsevier}
}

@book{hill1986introduction,
  title={An introduction to statistical thermodynamics},
  author={Hill, Terrell L},
  year={1986},
  publisher={Courier Corporation}
}

@article{neccsulea2007new,
  title={A new method for assessing the effect of replication on DNA base composition asymmetry},
  author={Nec{\c{s}}ulea, Anamaria and Lobry, Jean R},
  journal={Molecular biology and evolution},
  volume={24},
  number={10},
  pages={2169--2179},
  year={2007},
  publisher={Society for Molecular Biology and Evolution}
}

@article{parker2017mechanisms,
  title={Mechanisms and regulation of DNA replication initiation in eukaryotes},
  author={Parker, Matthew W and Botchan, Michael R and Berger, James M},
  journal={Critical reviews in biochemistry and molecular biology},
  volume={52},
  number={2},
  pages={107--144},
  year={2017},
  publisher={Taylor \& Francis}
}

@article{leonard2013dna,
  title={DNA replication origins},
  author={Leonard, Alan C and M{\'e}chali, Marcel},
  journal={Cold Spring Harbor perspectives in biology},
  volume={5},
  number={10},
  pages={a010116},
  year={2013},
  publisher={Cold Spring Harbor Lab}
}

@article{zhang2003z,
  title={The Z curve database: a graphic representation of genome sequences},
  author={Zhang, Chun-Ting and Zhang, Ren and Ou, Hong-Yu},
  journal={Bioinformatics},
  volume={19},
  number={5},
  pages={593--599},
  year={2003},
  publisher={Oxford University Press}
}

@Article{santalucia2004thermodynamics,
  author    = {SantaLucia Jr, John and Hicks, Donald},
  title     = {The thermodynamics of DNA structural motifs},
  journal   = {Annu. Rev. Biophys. Biomol. Struct.},
  year      = {2004},
  volume    = {33},
  pages     = {415--440},
  publisher = {Annual Reviews},
}

@article{pearson1996inverted,
  title={Inverted repeats, stem-loops, and cruciforms: significance for initiation of DNA replication},
  author={Pearson, Christopher E and Zorbas, Haralabos and Price, Gerald B and Zannis-Hadjopoulos, Maria},
  journal={Journal of cellular biochemistry},
  volume={63},
  number={1},
  pages={1--22},
  year={1996},
  publisher={Wiley Online Library}
}

@article{cheung2004palindrome,
  title={Palindrome regeneration by template strand-switching mechanism at the origin of DNA replication of porcine circovirus via the rolling-circle melting-pot replication model},
  author={Cheung, Andrew K},
  journal={Journal of virology},
  volume={78},
  number={17},
  pages={9016--9029},
  year={2004},
  publisher={Am Soc Microbiol}
}

@article{bikard2010folded,
  title={Folded DNA in action: hairpin formation and biological functions in prokaryotes},
  author={Bikard, David and Loot, C{\'e}line and Baharoglu, Zeynep and Mazel, Didier},
  journal={Microbiology and Molecular Biology Reviews},
  volume={74},
  number={4},
  pages={570--588},
  year={2010},
  publisher={Am Soc Microbiol}
}

@Article{sequence_dependent,
  author    = {Subramanian, Hemachander and Gatenby, Robert A},
  title     = {Evolutionary advantage of anti-parallel strand orientation of duplex DNA},
  journal   = {Scientific Reports},
  year      = {2020},
  volume    = {10},
  number    = {1},
  pages     = {9883},
  publisher = {Nature Publishing Group UK London},
}

@book{van1992stochastic,
  title={Stochastic processes in physics and chemistry},
  author={Van Kampen, Nicolaas Godfried},
  volume={1},
  year={1992},
  publisher={Elsevier}
}

@Article{symmetry_breaking,
  author    = {Subramanian, Hemachander and Gatenby, Robert A},
  title     = {Evolutionary advantage of directional symmetry breaking in self-replicating polymers},
  journal   = {Journal of theoretical biology},
  year      = {2018},
  volume    = {446},
  pages     = {128--136},
  publisher = {Elsevier},
}

@Book{inversemethod,
  author    = {Suhov, Yuri and Kelbert, Mark},
  title     = {Probability and statistics by example: volume 2, Markov chains: a primer in random processes and their applications},
  year      = {2008},
  volume    = {2},
  publisher = {Cambridge University Press},
}

@Article{master_equation,
  author    = {Berezhkovskii, Alexander M and Szabo, Attila},
  title     = {Committors, first-passage times, fluxes, Markov states, milestones, and all that},
  journal   = {The Journal of chemical physics},
  year      = {2019},
  volume    = {150},
  number    = {5},
  pages     = {054106},
  publisher = {AIP Publishing LLC},
}

@Article{inversemethod1,
  author  = {Zolaktaf, Sedigheh and Dannenberg, Frits and Schmidt, Mark and Condon, Anne and Winfree, Erik},
  title   = {The pathway elaboration method for mean first passage time estimation in large continuous-time Markov chains with applications to nucleic acid kinetics},
  journal = {arXiv preprint arXiv:2101.03657},
  year    = {2021},
}

@article{buchete2008coarse,
  title={Coarse master equations for peptide folding dynamics},
  author={Buchete, Nicolae-Viorel and Hummer, Gerhard},
  journal={The Journal of Physical Chemistry B},
  volume={112},
  number={19},
  pages={6057--6069},
  year={2008},
  publisher={ACS Publications}
}

@article{tillier2000contributions,
  title={The contributions of replication orientation, gene direction, and signal sequences to base-composition asymmetries in bacterial genomes},
  author={Tillier, Elisabeth RM and Collins, Richard A},
  journal={Journal of Molecular Evolution},
  volume={50},
  pages={249--257},
  year={2000},
  publisher={Springer}
}

@article{mclean1998base,
  title={Base composition skews, replication orientation, and gene orientation in 12 prokaryote genomes},
  author={McLean, Michael J and Wolfe, Kenneth H and Devine, Kevin M},
  journal={Journal of molecular evolution},
  volume={47},
  pages={691--696},
  year={1998},
  publisher={Springer}
}

@article{chen1992energy,
  title={Energy flow considerations and thermal fluctuational opening of DNA base pairs at a replicating fork: unwinding consistent with observed replication rates},
  author={Chen, YZ and Zhuang, W and Prohofsky, EW},
  journal={Journal of Biomolecular Structure and Dynamics},
  volume={10},
  number={2},
  pages={415--427},
  year={1992},
  publisher={Taylor \& Francis}
}

@article{rocha2004replication,
  title={The replication-related organization of bacterial genomes},
  author={Rocha, Eduardo PC},
  journal={Microbiology},
  volume={150},
  number={6},
  pages={1609--1627},
  year={2004},
  publisher={Microbiology Society}
}

@article{niu2003strand,
  title={Strand compositional asymmetries of nuclear DNA in eukaryotes},
  author={Niu, Deng K and Lin, Kui and Zhang, Da-Yong},
  journal={Journal of molecular evolution},
  volume={57},
  pages={325--334},
  year={2003},
  publisher={Springer}
}

@article{dai2005dna,
  title={DNA replication origins in the Schizosaccharomyces pombe genome},
  author={Dai, Jianli and Chuang, Ray-Yuan and Kelly, Thomas J},
  journal={Proceedings of the National Academy of Sciences},
  volume={102},
  number={2},
  pages={337--342},
  year={2005},
  publisher={National Acad Sciences}
}

@article{marsolier2012asymmetry,
  title={Asymmetry indices for analysis and prediction of replication origins in eukaryotic genomes},
  author={Marsolier-Kergoat, Marie-Claude},
  year={2012},
  publisher={Public Library of Science San Francisco, USA}
}

@article{zhang2005identification,
  title={Identification of replication origins in archaeal genomes based on the Z-curve method},
  author={Zhang, Ren and Zhang, Chun-Ting},
  journal={Archaea},
  volume={1},
  number={5},
  pages={335--346},
  year={2005},
  publisher={Wiley Online Library}
}

@article{dong2023doric,
  title={DoriC 12.0: an updated database of replication origins in both complete and draft prokaryotic genomes},
  author={Dong, Mei-Jing and Luo, Hao and Gao, Feng},
  journal={Nucleic Acids Research},
  volume={51},
  number={D1},
  pages={D117--D120},
  year={2023},
  publisher={Oxford University Press}
}

@article{sernova2008identification,
  title={Identification of replication origins in prokaryotic genomes},
  author={Sernova, Natalia V and Gelfand, Mikhail S},
  journal={Briefings in Bioinformatics},
  volume={9},
  number={5},
  pages={376--391},
  year={2008},
  publisher={Oxford University Press}
}

\end{document}